\documentclass[ reprint, aps, prl, two column, superscriptaddress, showkeys]{revtex4-1}
\usepackage{graphicx,epsfig}
\usepackage{amssymb}
\usepackage{amsmath}
\usepackage{bm}
\usepackage{textcomp}

\usepackage{color}

\newcommand* {\kk}{{\bm{k}}}
\newcommand* {\xd}{{[\bar{1}10]}}
\newcommand* {\yd}{{[110]}}
\newcommand* {\eps}{\epsilon}
\newcommand* {\frack}[2]{{\textstyle\frac{#1}{#2}}}

\begin{document}
\setcounter{page}{1}

\title[]{Transference of Fermi Contour Anisotropy to Composite Fermions}
\author{Insun \surname{Jo}}
\author{K. A. Villegas \surname{ Rosales}}
\author{M. A. \surname{Mueed}}
\author{L. N. \surname{Pfeiffer}}
\author{K. W. \surname{West}}
\author{K. W. \surname{Baldwin}}
\affiliation{Department of Electrical Engineering, Princeton University, Princeton, NJ 08544, USA  }
\author{R. \surname{Winkler}}
\affiliation{Department of Physics, Northern Illinois University, DeKalb, IL 60115, USA   }
\author{Medini \surname{Padmanabhan}}
\affiliation{Physical Sciences Department, Rhode Island College, Providence, RI 02908, USA  }
\author{M. \surname{Shayegan}}
\affiliation{Department of Electrical Engineering, Princeton University, Princeton, NJ 08544, USA  }
\date{\today}

\begin{abstract}

There has been a surge of recent interest in the role of anisotropy in interaction-induced phenomena in two-dimensional (2D) charged carrier systems. A fundamental question is how an anisotropy in the energy-band structure of the carriers at zero magnetic field affects the properties of the interacting particles at high fields, in particular of the composite fermions (CFs) and the fractional quantum Hall states (FQHSs). We demonstrate here tunable anisotropy for holes and hole-flux CFs confined to GaAs quantum wells, via applying \textit{in situ} in-plane strain and measuring their Fermi wavevector anisotropy through commensurability oscillations. For strains on the order of $10^{-4}$ we observe significant deformations of the shapes of the Fermi contours for both holes and CFs. The measured Fermi contour anisotropy for CFs at high magnetic field ($\alpha_\mathrm{CF}$) is less than the anisotropy of their low-field hole (fermion) counterparts ($\alpha_\mathrm{F}$), and closely follows the relation: $\alpha_\mathrm{CF} = \sqrt{\alpha_\mathrm{F}}$. The energy gap measured for the $\nu = 2/3$ FQHS, on the other hand, is nearly unaffected by the Fermi contour anisotropy up to $\alpha_\mathrm{F} \sim 3.3$, the highest anisotropy achieved in our experiments.
  
\end{abstract}
\maketitle

High-mobility, two-dimensional (2D), charged carriers at high perpendicular magnetic fields $B$ and low temperatures exhibit rich many-body physics driven by Coulomb interaction. Examples include the fractional quantum Hall state (FQHS), Wigner crystal, and stripe phase \cite{Shayegan.Review.2006, Jain.Book.2007}. Recently, the role of \textit{anisotropy} has become a focus of new studies \cite{Balagurov.PRB.2000, Shayegan.PSSB.2006, Gokmen.NatPhys.2010, Fradkin.ARCMP.2010, Xia.NatPhys.2011, Koduvayur.PRL.2011, Liu.PRB.2013, Haldane.PRL.2011, Kamburov.PRL.2013, Kamburov.PRB.2014, Mueed.PRL.2015a,Mueed.PRL.2015b,Mueed.PRB.2016, Mulligan.PRB.2010, Samkharadze.NatPhy.2016, Wang.PRB.2012, Qiu.PRB.2012, BYang.PRB.2012, KYang.PRB.2013, Balram.PRB.2016, Johri.NJP.2016}. This interest has been amplified by the recognition that, although the FQHSs at fillings $1/q$ ($q=$ odd integer) are well described by Laughlin's wave function with a rotational symmetry \cite{Laughlin.PRL.1983}, there is a geometric degree of freedom associated with the anisotropy of the 2D carrier system \cite{Haldane.PRL.2011}.

\begin{figure} [b!]
  \begin{center}
    \psfig{file=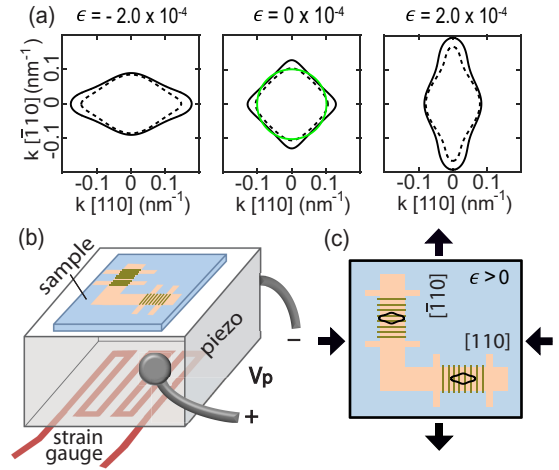, width=0.41\textwidth }
  \end{center}
  \caption{\label{fig1} (a) Calculated Fermi contours of GaAs holes at density $p = 1.8\times 10^{11}$~cm$^{-2}$ as a function of strain $\epsilon$ along the $[\bar{1}10]$ direction. Solid and dashed contours represent two spin-split subbands; the green circle with radius $k_{0} = \sqrt{2 \pi p}$ shows a spin-degenerate, circular Fermi contour at the same density. (b) Schematic of the experimental setup showing a thinned GaAs wafer glued on a piezo-actuator. A strain gauge mounted underneath measures the strain along $[\bar{1}10]$. (c) Sample fabricated to an L-shaped Hall bar has regions with electron-beam resist gratings on the surface. Thick arrows indicate the deformation of the crystal when a positive voltage $V_{P}$ is applied to the piezo. The resulting deformed cyclotron orbits are shown in black; note that these are rotated by $90^{\circ}$ with respect to the Fermi contours in reciprocal space. The shapes of the orbits and therefore the Fermi contours are determined via commensurability oscillations measurements.}
\end{figure}

\begin{figure*}[t!]
  \begin{center}
    \psfig{file=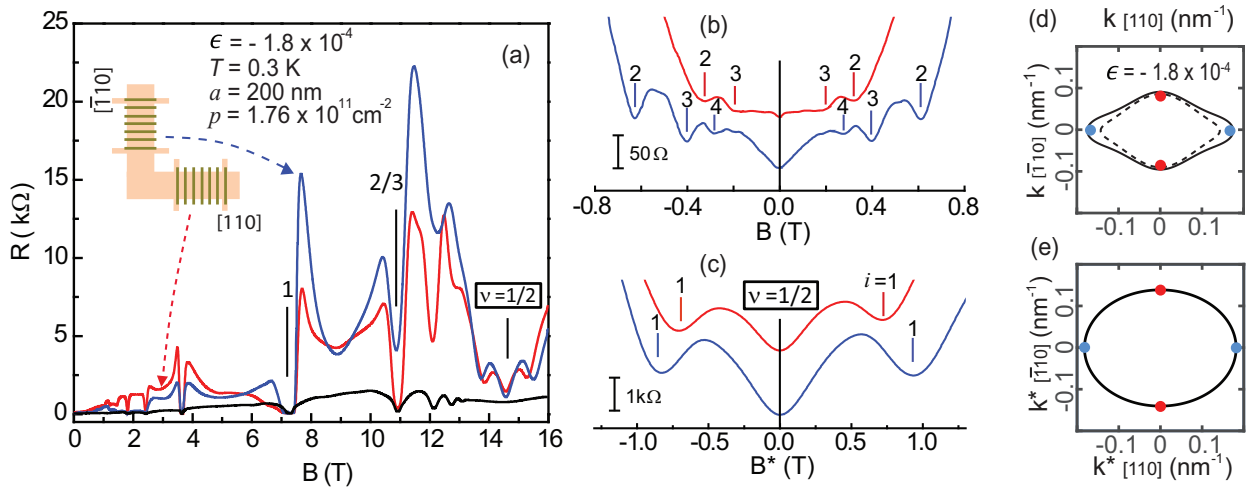, width=0.92\textwidth }
  \end{center}
  \caption{\label{fig2} (a) Magnetoresistance traces taken from different regions of the Hall bar when strain $\epsilon = -1.8\times 10^{-4}$ is applied along $[\bar{1}10]$. The red and blue traces are for the patterned regions along the $[110]$ and $[\bar{1}10]$ arms, while the black trace is for an unpatterned region. (b), (c) The red and blue traces in (a) are shown enlarged, exhibiting commensurability features for holes at low fields (b), and for CFs at high fields (c) near Landau level filling factor $\nu = 1/2$. The effective field for CFs in (c) is shown as $B^\ast$. The vertical lines in (b) and (c) indicate the positions of minima satisfying the commensurability conditions for holes and CFs (see text). (d) Calculated Fermi contours of spin-split holes at $p = 1.8 \times 10^{11}$~cm$^{-2}$ and $\epsilon = -1.8\times 10^{-4}$. Red and blue dots represent the measured Fermi wavevectors along the $[\bar{1}10]$ and $[110]$ directions, using traces in (b). (e) An elliptical Fermi contour for CFs based on the measured Fermi wavevectors, using traces in (c). }
\end{figure*}

The fundamental issue we address here is how the anisotropy of the energy-band structure of the low-field carriers transfers to the interacting particles at high $B$ and, in particular, to the FQHSs and composite fermions (CFs). The latter are electron-flux quasi-particles that form a Fermi sea at a half-filled Landau level \cite{Jain.Book.2007, Halperin.PRB.1993}, and provide a simple explanation for the nearby FQHSs \cite{Jain.PRL.1989}. There is no clear theoretical verdict yet. While some theories predict that the CF Fermi contour anisotropy ($\alpha_\mathrm{CF}$) should be the same as the zero-field (fermion) contour anisotropy ($\alpha_\mathrm{F}$) \cite{Balagurov.PRB.2000, Balram.PRB.2016}, others conclude that $\alpha_\mathrm{CF}$ is noticeably smaller than $\alpha_\mathrm{F}$ \cite{BYang.PRB.2012, KYang.PRB.2013, Johri.NJP.2016}. This question was also addressed in several recent experimental studies. For 2D electrons occupying AlAs conduction-band valleys with an anisotropic effective mass, a pronounced \textit{transport} anisotropy was reported for CFs, but the anisotropy of the CF \textit{Fermi contour} could not be measured because of the insufficient sample quality \cite{Gokmen.NatPhys.2010}. More recently, experiments probed the Fermi contour anisotropy of low-field carriers (both electrons and holes), and of CFs in GaAs quantum wells by subjecting them to an additional parallel magnetic field ($B_\|$) \cite{Kamburov.PRL.2013, Kamburov.PRB.2014, Mueed.PRL.2015a, Mueed.PRL.2015b, Mueed.PRB.2016}. However, the $B_\|$-induced anisotropy is primarily caused by the coupling between the in-plane and out-of-plane motions of the carriers, rendering a theoretical understanding of the data challenging. Furthermore, a strong $B_\|$ can lead to a bilayer-like charge distribution \cite{Mueed.PRL.2015b}.

As highlighted in Fig.~1, we demonstrate a simple yet powerful technique to tune and probe the anisotropy of both low-field carriers and high-field CFs \textit{without} applying $B_\|$. The experiments consist of subjecting the sample, a GaAs 2D hole system (2DHS), to strain \cite{fnote1, Shayegan.APL.2003, Shkolnikov.APL.2004,Supple} and measuring $\alpha_\mathrm{F}$ and $\alpha_\mathrm{CF}$, via commensurability oscillations measurements. We find that, for a given value of strain, CFs are less anisotropic than their low-field 2D hole counterparts, and the anisotropies are related through a simple empirical relation: $\alpha_\mathrm{CF} = \sqrt{\alpha_\mathrm{F}}$. In contrast, the measured energy gap of the $\nu = 2/3$ FQHS remains almost constant even for $\alpha_\mathrm{F}$ as large as 3.3. Our results allow a direct and quantitative comparison with theoretical predictions.

Figure 1(a) shows the results of numerical calculations for the strain-induced Fermi contour anisotropy of our sample, a 2DHS confined to a 175-{{\AA}}-wide GaAs (001) quantum well \cite{fnote2, Kamburov.PRB.2012, Kamburov.PRL.2012}. The self-consistent calculations are based on an $8\times8$ Kane Hamiltonian \cite{Supple,Bir.Book,Winkler.Book}. Without strain ($\epsilon = 0$), the Fermi contour of holes is four-fold symmetric but is split into two contours because of the spin-orbit interaction \cite{Winkler.Book}. The minority-spin contour is nearly circular while the majority-spin contour is warped. When tensile strain ($\epsilon > 0$) is applied along $[\bar{1}10]$, the hole Fermi contours become elongated along $[\bar{1}10]$ and shrink along the $[110]$ direction \cite{Supple,Bir.Book,Winkler.Book, Kolokolov.PRB.1999, Habib.PRB.2007, Shabani.PRL.2008}. On the other hand, compressible strain ($\epsilon < 0$) has the opposite effect [Fig.\ 1(a)]. Our experimental setup for applying \textit{in situ} tunable strain to the sample is shown in Figs. 1(b) and 1(c) \cite{Shayegan.APL.2003}. An L-shaped Hall bar is etched into the GaAs wafer which is thinned to $\sim$120 $\mu$m and glued on one surface of a stacked piezo-actuator. When a voltage $V_{P} > 0$ ($V_{P} < 0$) is applied to the piezo, the sample expands (contracts) along $[\bar{1}10]$. This is monitored using a strain gauge glued to the opposite face of the piezo \cite{Shayegan.APL.2003,Shkolnikov.APL.2004,Supple}.

\begin{figure*}
  \begin{center}
    \psfig{ file=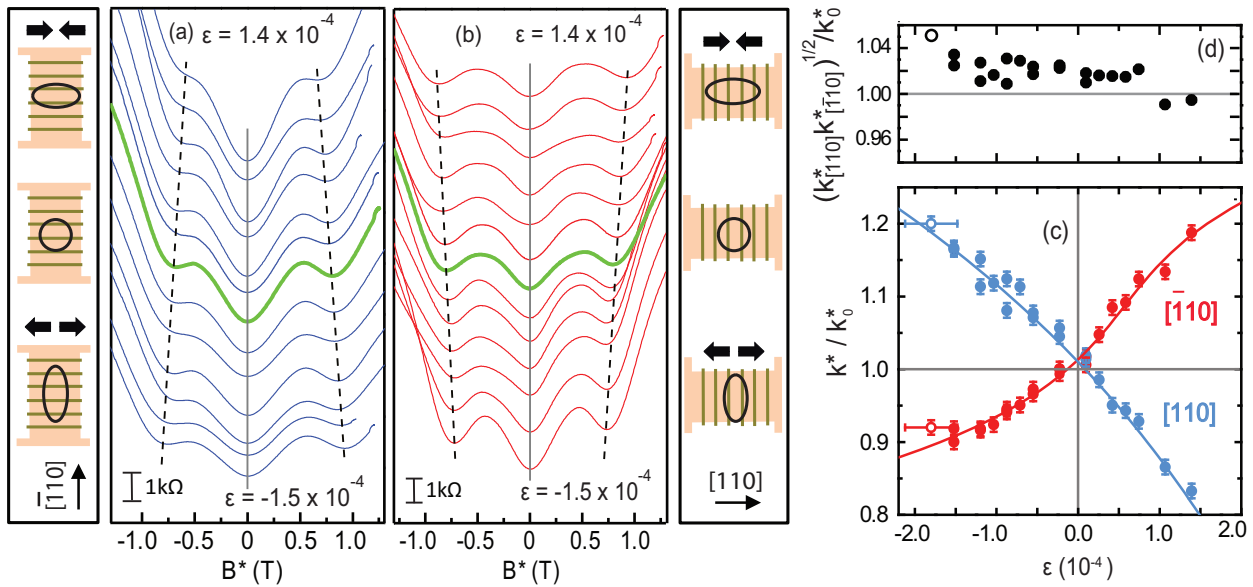, width=0.92\textwidth }
  \end{center}
  \caption{\label{fig3} (a) and (b) Commensurability features for CFs near $\nu = 1/2$ along the $[\bar{1}10]$ and $[110]$ directions of the Hall bar as strain $\epsilon$ is varied between $-1.5$ and $+1.4 \times 10^{-4}$. Green traces are for the $\epsilon = 0$ case. Dashed lines are guides to the eye to follow the evolution of the CF commensurability minima. Left panel of (a) and right panel of (b) show the direction of the strain (thick arrows), and shapes of CF cyclotron orbits (circle and ellipses). (c) Measured CF Fermi wavevector $k^\ast$ along the $[\bar{1}10]$ (red) and $[110]$ (blue) directions, normalized to $k^\ast_{0}$, are shown as a function of $\epsilon$. The lines are least squares fits to the measured data and serve as guides to the eye. Open circles are from a different sample cool-down and represent the data shown in Fig.~2. (d) Geometric means of $k^\ast_{[110]}$ and $k^\ast_{[\bar{1}10]}$, normalized to $k^\ast_{0}$.}
\end{figure*}

In order to measure the Fermi wavevectors, we fabricate periodic gratings of negative electron-beam resist, with period $a = 200$ nm, on the surface of the L-shaped Hall bar [Fig.\ 1(c)]. The grating induces a periodic strain onto the GaAs surface which in turn results in a small periodic modulation of the 2DHS density via the piezoelectric effect \cite{Endo.PRB.2000, Kamburov.PRB.2012, Kamburov.PRL.2012}. In the presence of $B$, when the cyclotron motion of holes becomes commensurate with $a$, the magnetoresistance shows oscillations whose minima positions are directly related to the carriers' Fermi wavevector in the direction perpendicular to the current \cite{Endo.PRB.2000, Kamburov.PRB.2012, Kamburov.PRL.2013, Gunawan.PRL.2004}. Figure 1(c) shows an example when tensile strain is applied along $[\bar{1}10]$; the elongated cyclotron orbits under a finite $B$ are indicated by black curves.

Figure 2(a) shows magnetoresistance traces for $\epsilon = -1.8 \times 10^{-4}$. The red and blue traces are from the patterned regions along the $[110]$ and $[\bar{1}10]$ directions, respectively, while the black trace is for an unpatterned region. The red and blue traces exhibit commensurability features for holes near $B = 0$ [Fig.\ 2(b)] and for CFs near $\nu = 1/2$ [Fig.\ 2(c)]. To analyze the low-field hole data, we use the \textit{electrostatic} commensurability condition \cite{Weiss.EPL.1989,Winkler.PRL.1989,Gerhardts.PRL.1989,Beenakker.PRL.1989,Endo.PRB.2000,Kamburov.PRB.2012} for the minima positions, $2R_{c}/a = i-1/4$ ($i = 1,2,3, \ldots$) where 2$R_{c} = 2\hbar k/eB$ is the cyclotron orbit diameter, $k$ is the 2DHS Fermi wavevector perpendicular to the current, and $a$ is the period of the density modulation. For CFs, we observe commensurability features near $\nu = 1/2$, or $B_{1/2} = 14.5$~T [Fig.\ 2(c)]. The positions of minima around $\nu = 1/2$ yield the Fermi wavevector of CFs ($k^\ast$) according to the \textit{magnetic} commensurability condition \cite{Smet.PRL.1999,Zwerschke.PRL.1999,Kamburov.PRL.2012}, $2R^\ast_{c}/a = i + 1/4$, where the CF cyclotron diameter 2$R^\ast_{c} = 2\hbar k^\ast / eB^\ast$ and $B^\ast = B-B_{1/2}$ is the effective field for CFs \cite{Kamburov.PRL.2014,fnote3}. 

In Fig.\ 2(d) we mark the measured Fermi wavevectors for holes with red and blue dots along $[\bar{1}10]$ and $[110]$. Although theoretical calculations for holes predict two spin subbands with different Fermi wavevectors [black solid and dashed curves in Fig.\ 2(d)], we measure a single $k$ for each direction from the commensurability features. The measured $k_{[\bar{1}10]}$ (red dots) and $k_{[110]}$ (blue dots) are close to the average calculated Fermi wavevectors for the two spin-subbands. Figure 2(e) shows $k^\ast$ measured for CFs with red and blues dots. We depict the Fermi contour as an ellipse because there are no theoretical calculations available for CFs, and also the area of an ellipse spanned by the two measured $k^\ast$ accounts for the density of CFs which are fully spin-polarized at high fields \cite{Kamburov.PRL.2012}. Note that the CF Fermi contour anisotropy $\alpha_\mathrm{CF} \equiv k^\ast_{[\bar{1}10]} / k^\ast_{[110]}= 0.77$, which is closer to unity than the 2D hole anisotropy $\alpha_\mathrm{F} \equiv k_{[\bar{1}10]} / k_{[110]} = 0.53$. Quantitatively, we find $\alpha_\mathrm{CF} = \sqrt{\alpha_\mathrm{F}}$ to within 5\%; see below.

Next we demonstrate the tunability of CF Fermi contour anisotropy via strain. Figures 3(a) and 3(b) show magnetoresistance traces near $\nu = 1/2$, taken along $[\bar{1}10]$ and $[110]$, at different strains. In each panel, the green trace represents the $\epsilon = 0$ case where the Fermi contour is essentially isotropic and $k^\ast = k^\ast_{0} = \sqrt{4 \pi p}$ \cite{Supple}. The traces shown above the green trace are for tensile strain ($\epsilon>0$) while those below are for compressive strain ($\epsilon<0$). In Fig.\ 3(a) the positions of resistance minima move towards (away from) $B^\ast = 0$ for $\epsilon>0$ ($\epsilon<0$), while the opposite is true for Fig.\ 3(b). These observations imply a distortion in the shape of CF cyclotron orbits as depicted in the side panels of Figs.\ 3(a) and 3(b).

\begin{figure}
  \begin{center}
    \psfig{file=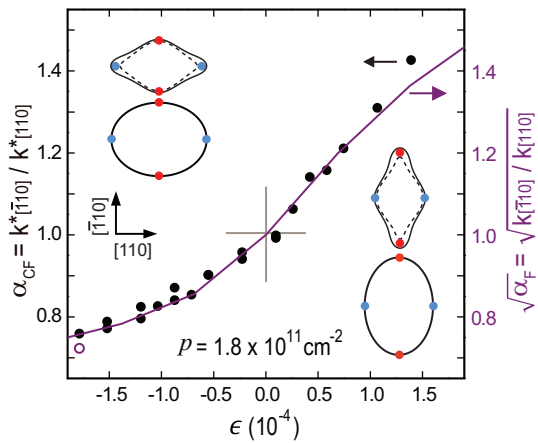, width=0.4\textwidth }
  \end{center}
  \caption{\label{fig4} Strain-dependent Fermi contour anisotropy of holes ($\alpha_\mathrm{F}$) and CFs ($\alpha_\mathrm{CF}$). Black circles are the measured $\alpha_\mathrm{CF}$. The purple curve represents the square-root of the calculated $\alpha_\mathrm{F}$. The open purple circle shows the measured $\alpha_\mathrm{F}$ for holes as described in Fig.~2. The left and right insets show the hole and CF Fermi contour shapes for $\epsilon= -1.5$ and $1.4 \times 10^{-4}$.}
\end{figure}

Figure 3(c) summarizes the measured $k^\ast$ along $[\bar{1}10]$ and $[110]$, normalized by $k^\ast_{0}$. Comparing $k^\ast$ values for compressive and tensile cases, the change of $k^\ast$ for $\epsilon > 0$ is larger than for $\epsilon < 0$. This asymmetry reflects the response of the 2DHS Fermi contour to the applied strain [Fig.\ 1(a)]. We also find that the geometric means of $k^\ast / k^\ast_{0}$ along the two perpendicular directions remain close to unity [Fig.\ 3(d)]. This suggests that CF Fermi contours are nearly elliptical, although we cannot exclude a more complex shape.

Figure 4 illustrates the highlight of our study: comparison of strain-induced Fermi contour anisotropy for CFs and holes. The measured anisotropy for CFs, $\alpha_\mathrm{CF} = k^\ast_{[\bar{1}10]} / k^\ast_{[110]}$, is shown by black circles, and the \textit{square-root} of the calculated anisotropy for holes, $\alpha_\mathrm{F} = k_{[\bar{1}10]} / k_{[110]}$, by a purple curve. Here we use, for each $k_{[\bar{1}10]}$ and $k_{[110]}$, the averaged values of $k$ for the spin-subbands, since experiments measure only a single $k$ for each direction. Remarkably, the measured $\alpha_\mathrm{CF}$ for CFs essentially coincides with $\sqrt{\alpha_\mathrm{F}}$ over the entire range of strains applied in the experiments. This is particularly striking because there are no fitting or adjustable parameters.

Lastly, we study the impact of anisotropy on the strength of FQHSs, focusing on the energy gap for the $\nu = 2/3$ state. The sample used for the measurements has $p = 1.3 \times 10^{11}$~cm$^{-2}$, and exhibits commensurability features only for holes along $k_{[\bar{1}10]}$. Moreover, using a different cool-down procedure \cite{Supple}, we achieved larger strain values ($\epsilon$ up to $5.5 \times 10^{-4}$), and anisotropy ($\alpha_\mathrm{F}$ as large as 3.3) as shown in Fig.~5. The measured energy gap $\Delta$, determined from the expression $R(T) \sim e^{-\Delta / 2 T}$, is 2.1~K for $\epsilon = 0$, and it decreases only to 2.0~K even for a large anisotropy $\alpha_\mathrm{F} = 3.3$. The small decrease of $\Delta$ is consistent with recent theoretical predictions \cite{Balram.PRB.2016}, suggesting that the FQHSs in the lowest Landau level are quite robust against anisotropy.

\begin{figure}[t!]
  \begin{center}
    \psfig{file=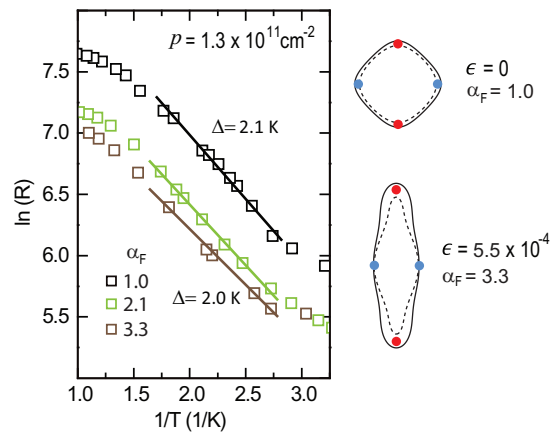, width=0.4\textwidth }
  \end{center}
  \caption{\label{fig5} (a) Longitudinal resistance at $\nu = 2/3$ is recorded at different temperatures for energy gap ($\Delta$) measurements, in cases of $\alpha_\mathrm{F} = 1.0$, 2.1, and 3.3. The slopes of the straight lines yield the energy gaps. The calculated 2DHS Fermi contour shapes are shown on the right, where the blue dots indicate the measured $k_{[\bar{1}10]}$ and red dots are determined based on calculations for $p = 1.3 \times 10^{11}$~cm$^{-2}$. }
\end{figure}

Returning to the Fermi contour anisotropy, our measurements (Fig.~4) provide quantitative evidence for a simple relation between the anisotropy of low-field fermions and high-field CFs: $\alpha_\mathrm{CF} = \sqrt{\alpha_\mathrm{F}}$. This appears to contradict some of the theories which predict that $\alpha_\mathrm{F}$ and $\alpha_\mathrm{CF}$ should be the same \cite{Balagurov.PRB.2000, Balram.PRB.2016}. One can, however, qualitatively justify the square-root relation \cite{Gokmen.NatPhys.2010}. In an ideal, isotropic 2D system, the Coulomb interaction $V_\mathrm{C}$ ($\propto 1/\sqrt{x^2 + y^2}$) determines the physical parameters of CFs, including their effective mass $m^\ast$ which is linearly proportional to $V_\mathrm{C}$ \cite{Jain.Book.2007, Halperin.PRB.1993}. At a given filling factor, $V_\mathrm{C}$ is quantified solely by the magnetic length $l_{B} = \sqrt{\hbar/eB}$ \cite{Jain.Book.2007,Halperin.PRB.1993}. Now, a system with an anisotropic dispersion $\alpha_\mathrm{F} \ne 1$ at $B = 0$ can be mapped to a system with an isotropic Fermi contour and an anisotropic $V_\mathrm{C}$ ($\propto 1/\sqrt{x^{2}\alpha_\mathrm{F} + y^{2}/\alpha_\mathrm{F}}$) using the coordinate transformations $x \rightarrow x/\sqrt{\alpha_\mathrm{F}}$ and $y \rightarrow y \sqrt{\alpha_\mathrm{F}}$. In such a system the strength of $V_\mathrm{C}$ at high fields thus depends not only on $l_B$ but also on the direction, i.e., $V_\mathrm{C}$ anisotropy is $\alpha_\mathrm{F}$. If one assumes that CFs have a parabolic dispersion and an anisotropic $m^\ast $ whose anisotropy follows linearly the anisotropy of $V_\mathrm{C}$, then the \textit{mass} anisotropy of CFs is given by $\alpha_\mathrm{F}$, implying that their Fermi wavevector anisotropy is proportional to $\sqrt{\alpha_\mathrm{F}}$.

In conclusion, our results provide direct and quantitative evidence for the inheritance of Fermi contour anisotropy by CFs from their low-field fermion counterparts through a simple relation: $\alpha_\mathrm{CF} = \sqrt{\alpha_\mathrm{F}}$. While the discussion in the preceding paragraph serves as a plausibility argument for this relation, there is also some very recent rigorous theoretical justification. In their numerical calculations for anisotropic fermions with a parabolic band, Ippoliti \textit{et al.} \cite{Ippoliti.2017} find that the relation $\alpha_\mathrm{CF} = \sqrt{\alpha_\mathrm{F}}$ is indeed empirically obeyed \cite{fnote4}. It remains to be seen, both experimentally and theoretically, if the relation holds when the fermions' band deviates significantly from parabolic \cite{fnote5,Mueed.PRL.2015a}.

\newpage
\appendix

\renewcommand{\thesection}{\Roman{section}}    
\renewcommand{\thefigure}{A\arabic{figure}}

\section{Supplemental Material: Transference of Fermi Contour Anisotropy to Composite Fermions}

\section{S I. Inclusion of strain in Fermi contour calculations}

The in-plane components of the strain created by the piezo-actuator
are transfered to the Hall bar, giving rise to the strain components
$\eps_\xd$ and $\eps_\yd$.  The effect of strain on the electronic
structure is usually expressed in a conventional coordinate system
with $\hat{\bm{x}} \parallel [100]$, $\hat{\bm{y}} \parallel [010]$,
and $\hat{\bm{z}} \parallel [001]$ (e.g., Refs.\
\cite{Bir.Book, Winkler.Book}).  Here the nonzero components
$\eps^c_{ij}$ of the second-rank strain tensor become
\begin{subequations}
  \begin{align}
    \eps^c_{xx} & = \eps_{yy}^c = \frack{1}{2} (\eps_\xd + \eps_\yd) \\
    \eps^c_{xy} & = \eps_{yx}^c = \frack{1}{2} (\eps_\xd - \eps_\yd)
    \equiv \frack{1}{2} \, \eps \label{eq:strain:xy} \\
    \eps^c_{zz} & = - \frac{c_{12}} {c_{11}} (\eps_\xd + \eps_\yd).
    \label{eq:strain:zz}
  \end{align}
\end{subequations}
Equation (\ref{eq:strain:xy}) defines the strain $\eps$ used in the
main paper.  It is this shear strain component that is primarily
responsible for the Fermi contour anisotropy
$\alpha_\mathrm{F} \ne 1$ at $B=0$.  Equation (\ref{eq:strain:zz})
follows from Hooke's law, assuming that the Hall bar's extent in $z$
direction can adjust freely to the in-plane strain.  The
coefficients $c_{11}$ and $c_{12}$ are the elastic constants of
GaAs.

We incorporate the effect of the strain $\eps^c_{ij}$ into our
self-consistent calculations using the Bir-Pikus strain Hamiltonian
\cite{Bir.Book, Winkler.Book}.  Numeric values for the relevant
deformation potentials are given in Ref.~\cite{Winkler.Book}.
We note that a simple analytical model can be developed that
evaluates the effect of strain using perturbation theory in lowest
order of the wave vector $\kk$ and strain $\eps^c_{ij}$ (similar to
the analytical models developed in Ref.~\cite{Winkler.Book}).
However, for the 2D hole systems studied here, terms of higher order
in $\kk$ and $\eps^c_{ij}$ are very important.  These terms are
fully taken into account in our numerical calculations.  The
lowest-order analytical model overestimates the 2D hole Fermi contour
anisotropy $\alpha_\mathrm{F}$ by about a factor of five.

\section{S II. Determination of strain in experiments}

Because of the different thermal contractions of the piezo-actuator and the sample, a residual, or ``built-in" strain ($\epsilon_\mathrm{bi}$) develops when the sample is cooled to low temperatures. The value of $\epsilon_\mathrm{bi}$ depends on the piezo-actuator and the details of the cool-down, and is not fully controllable. For example, when the two leads of the piezo-actuator are left open-circuited during the cool-down, $\epsilon_\mathrm{bi}$ is typically small. This is the case for the data shown in Fig. 3, where $\epsilon_\mathrm{bi} \simeq -0.5\times10^{-4}$. In presenting the data in Figs. 3 and 4, we have corrected for this $\epsilon_\mathrm{bi}$. To determine the value of $\epsilon_\mathrm{bi}$ and make the correction, we first take data at different values of biases ($V_\mathrm{P}$) applied to the piezo-actuator, and find the bias that gives $k_{[\bar{1}10]} =k_{[110]}$. This bias corresponds to $\epsilon=0$. We then determine the applied $\epsilon$ from the relative change of strain as monitored by the strain gauge glued to the piezo-actuator \cite{Shayegan.APL.2003,Shkolnikov.APL.2004,Gunawan.PRL.2004}. 

For the data of Fig. 5, the sample was cooled down multiple times with a 1 G$\Omega$ resistor across the two leads of the piezo-actuator. This led to large values of $\epsilon_\mathrm{bi}$, up to $5.5 \times 10^{-4}$ \cite{Shabani.PRL.2008}, and $\alpha_\mathrm{F}$ as large as 3.3, as shown in Fig.~5. Unfortunately, we could not achieve such large $\epsilon_\mathrm{bi}$ with the sample of Fig. 3.

\begin{acknowledgments}
  We acknowledge support by the DOE BES (DE-FG02-00-ER45841) grant
  for measurements, and the NSF (Grants DMR 1305691, DMR 1310199, MRSEC DMR 1420541, and ECCS 1508925), the Gordon and Betty Moore Foundation (Grant GBMF4420), and Keck Foundation for sample fabrication and characterization. We thank R. N. Bhatt, S. D. Geraedts, M. Ippoliti,  J. K. Jain, and D. Kamburov for illuminating discussions.

\end{acknowledgments}

\end{document}